\documentclass[10pt,letterpaper,compsoc,conference]{IEEEtran}
\IEEEoverridecommandlockouts

\usepackage{cite}
\usepackage{amsmath,amssymb,amsfonts}
\usepackage{algorithmic}
\usepackage{graphicx}
\usepackage[dvipsnames]{xcolor}
\usepackage[final]{microtype}
\usepackage[italic]{mathastext}
\usepackage{libertine}
\usepackage[T1]{fontenc}
\usepackage{textcomp}
\usepackage{url}
\usepackage[varqu,varl]{zi4}
\usepackage[all]{nowidow}
\usepackage[keeplastbox]{flushend}
\usepackage{accsupp}

\usepackage{booktabs}
\usepackage{tabularx}
\usepackage{multirow}
\usepackage{float}
\usepackage[section]{placeins}
\graphicspath{{figures/}{figs/}}

\usepackage{xcolor}

\begin{document}

\title{RAGMark: A Comprehensive Framework for Benchmarking Retrieval-Augmented Generation Systems
}

\author{
\IEEEauthorblockN{
Zlatan Feric\IEEEauthorrefmark{1},
Amir Taherin\IEEEauthorrefmark{1},
Bin Ren\IEEEauthorrefmark{3},
Yanzhi Wang\IEEEauthorrefmark{1},
Jennifer Dy\IEEEauthorrefmark{1},
David Kaeli\IEEEauthorrefmark{1}
}
\IEEEauthorblockA{
\IEEEauthorrefmark{1}Northeastern University, Boston, MA, USA \quad
\IEEEauthorrefmark{3}College of William \& Mary, VA, USA\\
\{feric.z, taherin.a, yanzhiwang, j.dy, d.kaeli\}@northeastern.edu \\
bren@cs.wm.edu}
}
\maketitle

\begin{abstract}
We present RAGMark, a modular benchmarking framework for advanced Retrieval-Augmented Generation (RAG) systems targeting small-scale multi-GPU environments. RAGMark evaluates diverse RAG components, including retrievers, vector databases, prompt-processing methods, and generator models, while collecting detailed per-stage metrics such as latency, GPU utilization, memory consumption, power usage, time to first token (TTFT), throughput, and answer quality. The framework is highly extensible, separating RAG stages, timing, and resource monitoring into modular components, and is designed to efficiently sweep large configuration spaces while minimizing repeated model and database initialization overhead.

Using RAGMark, we characterize five RAG workloads on open-domain QA datasets across varying retrieval depths, model scales, reranking, compression methods, and vector database configurations. We show that while autoregressive generation dominates latency in naive pipelines, context-reduction techniques shift bottlenecks across compute, memory bandwidth, and preprocessing stages. Reranking and compression produce compounding benefits: reranking reduces compression workload itself, while both jointly reduce prefill and KV-cache traversal costs, lowering energy consumption by up to 66\%. We further observe strong cross-stage interactions, where small upstream context reductions cascade through downstream latency, memory traffic, and energy consumption. The RAGMark source code is publicly available at: https://github.com/zferic/RAGMark.
\end{abstract}


\section{Introduction}

Large Language Models (LLMs), such as GPT-4~\cite{OpenAI2023GPT4}, LLaMA3~\cite{dubey2024llama}, and Mistral~\cite{Jiang2023Mistral7B}, have achieved impressive results across a diverse set of Natural Language Processing (NLP) tasks, including question answering, summarization, and code generation. Despite these advances, LLMs continue to face fundamental challenges, including hallucinations, outdated world knowledge, and restricted access to domain-specific knowledge. Retrieval-Augmented Generation (RAG)~\cite{lewis2020retrieval} has emerged as a practical solution by extending LLM capabilities through an information-retrieval component that accesses an external knowledge source and augments prompts with relevant passages to improve generation quality. As RAG pipelines grow more sophisticated, understanding their computational demands has become increasingly important, particularly in locally deployed environments where data sensitivity and resource constraints are significant concerns~\cite{seemakhupt2024edgerag,yu2025ragdoll}.

Even in a simple or naive RAG implementation, consisting of only retrieval and generation, system performance is shaped by a large number of parameters, including input query length, encoding model architecture and size~\cite{devlin2019bert, wolf2020transformers}, database index configuration~\cite{johnson2019billion, faiss_github}, database size, retrieval depth, final context length, generator model architecture~\cite{vaswani2017attention}, and output length. Furthermore, the allocation of resources across different RAG components can shift the computational bottleneck. For example, retrieval performed on a CPU versus a GPU can drastically change the latency characteristics. While most large-scale retrieval systems rely on CPU-resident indexes due to memory constraints, moderately sized indexes can be hosted on GPUs, resulting in a significant latency reduction, especially in long context or batch search scenarios. However, transferring indexes to GPU memory introduces additional memory overhead and may leave accelerator resources underutilized after retrieval completes.

The dominant bottleneck in a RAG pipeline is therefore highly configuration dependent. Prior work suggests that LLM inference is often the primary bottleneck in RAG systems~\cite{jin2024ragcache}, while others demonstrate that bottlenecks shift depending on retrieval architecture, database placement, and model scale~\cite{jiang2025rago}. For example, a large CPU-resident retrieval index paired with a small generator may become retrieval-bound, whereas a GPU-resident index paired with a large generator model (e.g., Llama 70B~\cite{grattafiori2024llama}) will typically be generation-bound. In addition, many practical RAG workloads involve large textual queries or multi-hop reasoning tasks, increasing computational demand before retrieval and generation even begin.

Modern RAG pipelines have evolved beyond the classical retrieve-then-generate architecture to include additional stages such as query rewriting~\cite{ma2023query,wang2023query2doc}, document reranking~\cite{nogueira2019passage, nogueira2020document}, and prompt compression~\cite{jiang2023llmlingua,pan2024llmlingua, li2023compressing}. These additional stages often rely on auxiliary encoder models or autoregressive LLMs, introducing additional inference and preprocessing overhead, while only modestly improving answer quality~\cite{gao2023retrieval,Gupta2024RAGsurvey}. This creates a fundamental tension between efficiency and quality, making RAG a uniquely challenging workload to characterize and optimize. 

Despite the growing complexity and widespread deployment of RAG systems, there is currently a lack of systematic benchmarking frameworks capable of capturing the full RAG configuration space, profiling individual pipeline stages, and analyzing resulting system-level trade-offs~\cite{gao2023retrieval,Gupta2024RAGsurvey}. Existing studies either focus narrowly on LLM inference~\cite{kwon2023vllm, jin2024ragcache} or treat RAG as a monolithic workload~\cite{jiang2025rago, shen2025hermes}, overlooking the rich interactions between embedding, retrieval, compression, reranking, and generation stages.

RAG workloads present several unique challenges for workload characterization:

\begin{itemize}
    \item \textbf{Heterogeneous Components:} RAG systems combine multiple distinct subsystems, including embedding models, vector databases, rerankers, compressors, and autoregressive generators, each with different individual configurations and distinct computational and memory characteristics.
    
    \item \textbf{Complex Performance Interactions:} Pipeline stages interact in non-trivial ways. For example, adding a reranking or compression method will reduce context length, but the additional computational overhead simultaneously increases pre-processing overhead latency and resource consumption, while either increasing or decreasing latency. 
    
    \item \textbf{Dynamic Query-Dependent Behavior:} Retrieval recall, reranking, compression effectiveness, and generation cost vary significantly across datasets, individual query complexity, and retrieved context length. Therefore, RAG exhibits different system performance across queries, so a RAG benchmark needs to adapt accordingly. 
    
    \item \textbf{Large Configuration Space:} Exploring combinations of retriever models, index types, compression methods, retrieval depths, and generator scales creates an extremely large benchmarking space that is expensive to evaluate exhaustively. 
\end{itemize}

To address these challenges, we present RAGMark, a user-friendly benchmarking and analysis framework targeted at studying small-scale multi-GPU environments. Our framework is designed to efficiently explore an extensive set of RAG configurations, while minimizing redundant computations. For example, vector databases and models are loaded once and reused across multiple experimental configurations, allowing large parameter sweeps without repeatedly rebuilding indexes or reloading models. The framework also supports dataset sampling strategies to reduce benchmarking time while preserving representative workload behavior. In addition, the framework produces detailed query-level and temporal trace logs, enabling both aggregate analysis and fine-grained per-stage profiling.

The key contributions of this work include:

\begin{itemize}
    \item We present RAGMark, a modular and extensible benchmarking framework for end-to-end RAG characterization, supporting configurable retrievers~\cite{hu2025kalmembedding, intfloatE5smallV2, intfloatE5baseV2, intfloatE5largeV2}, vector stores~\cite{faiss_github, johnson2019billion}, reranking methods, prompt compression techniques~\cite{jiang2023llmlingua, pan2024llmlingua, li2023compressing}, and generator models~\cite{llama32_1b_instruct, llama32_3b_instruct, llama30_8b_instruct, meta2025llama3}.
    
    \item We develop a unified measurement infrastructure for per-query, per-stage characterization of RAG systems, including latency, energy consumption, GPU/CPU utilization, memory usage, and response quality.
    
     \item We enable systematic exploration of the RAG design space through
    configuration-driven execution, including configurable retrieval and
    reranking depths, compression rates, model configurations, vector-index
    configurations, pipeline configurations, and mapping of RAG pipeline
    stages across CPUs and GPUs.
        
    \item Using RAGMark, we identify and quantify cross-stage interactions across representative RAG pipelines, showing how retrieval depth, reranking, compression, model scale, and index placement shift system bottlenecks, and how upstream context reduction can propagate into downstream reductions in generation latency, memory traffic, and energy consumption.

\end{itemize}

\begin{figure}[t]
    \centering
    \includegraphics[
        width=\columnwidth,
        trim={.5 1cm 2.2cm 1.7cm},
        clip
    ]{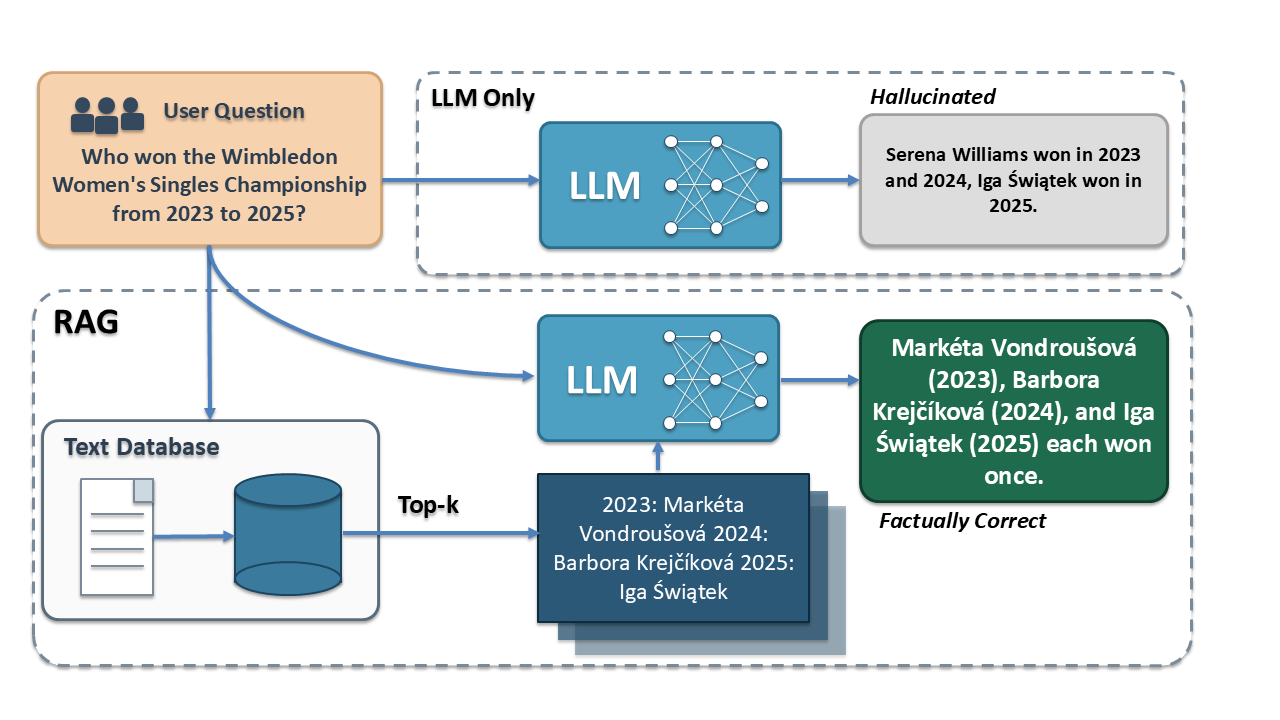}
    \caption{Workflow comparison: standalone LLM vs. RAG. The factually accurate answer is shown in green.}
    \label{fig:rag_simple}
\end{figure}
\section{Background}

\subsection{Retrieval-Augmented Generation}
Retrieval-Augmented Generation (RAG)~\cite{lewis2020retrieval} extends the capabilities of LLMs by incorporating an external information-retrieval component into the generation process. A simple two-stage RAG pipeline is shown in Figure~\ref{fig:rag_simple}, comparing it against LLM-only inference. Rather than relying solely on knowledge encoded in model parameters, RAG systems retrieve relevant passages from an external knowledge source at inference time and incorporate them into the prompt provided to the generator. This approach addresses several fundamental limitations of standalone LLMs, including hallucinations~\cite{huang2025survey}, outdated world knowledge~\cite{augenstein2024factuality}, and restricted access to domain-specific information~\cite{gao2023retrieval}.


\subsection{RAG Steps and Computational Workload}

While RAG computational requirements share similarities with LLM inference, RAG execution comes with additional computational overhead that distinguishes it from standard LLM inference. RAG is a heterogeneous workload that combines data-intensive retrieval stages with compute-intensive intermediate and generation stages, each with distinct compute requirements and performance characteristics.

\textbf{Encoding stage:} computes a dense vector representation of the user query using a small language model. These models are typically smaller encoder-based transformer architectures and requires matrix operations identical to those used in LLMs, without the autoregressive generation head. However, there are much larger encoders currently listed on HuggingFace, scaling up to billions of parameters. Furthermore, the encoding stage can consume more computational resources if the input document is large.

\textbf{Retrieval Stage:} uses a dense vector representation of the input to search over a large vector index. In the simplest approach, this is performed by computing the dot product, cosine distance (or other similarity metric) between the encoded query and all index vectors, then runs sorting to retrieve the top-$k$ closest matches (a GEMM matrix product), which are matched back to the raw text database using the index entries. The memory footprint of this index scales directly with corpus size, embedding dimensionality, and numerical precision --- storing $N$ vectors of dimension $d$ in FP16 requires $2Nd$ bytes, meaning a corpus of 9.2M chunks with 768-dimensional embeddings occupies roughly 14 GB, while larger embedding models (e.g., 1536 or 3072 dimensions) double or quadruple this footprint, respectively. For modestly sized corpora such as a Wikipedia dump~\cite{denoyer2006wikipedia}, this makes GPU placement a significant architectural decision. When the index fits in GPU memory, retrieval is fast (on the order of tens of milliseconds); when it must reside on the CPU, retrieval latency can increase by orders of magnitude~\cite{jiang2025rago}. The choice of index structure further affects this tradeoff, as discussed in Section~\ref{sec:vector_db}.

\textbf{Generation Stage:} consists of autoregressive LLM inference over the retrieved context and query, whose computational complexity scales quadratically with input length~\cite{vaswani2017attention}. LLM inference is typically decomposed into two stages: i) prefill, which processes the full input context and is compute-intensive, and ii) decode, which generates tokens autoregressively and is memory bandwidth-bound by using the KV-Cache. Beyond input length, the compute and memory requirements of LLM inference scale directly with model size. In terms of compute, a standard forward pass requires roughly 2 FLOPs per parameter per token. Consequently, a 7B model requires about 14 billion operations per token, while a 70B model demands roughly 140 billion operations. As a rule of thumb, storing model parameters in 8-bit precision requires roughly 1 GB per billion parameters --- a 7B model requires approximately 7 GB, while a 70B model requires roughly 70 GB of GPU memory, before accounting for activations and the KV cache. 

In a RAG system, the dominant bottleneck is therefore highly configuration dependent. Prior work has shown that LLM inference tends to dominate when the retrieval index is small~\cite{jin2024ragcache}, while retrieval becomes the bottleneck when the index is large or CPU-bound~\cite{jiang2025rago}. Factors such as retrieval depth (top-$k$), chunk size, embedding model size, and generator model scale all influence where the bottleneck lies. In advanced RAG pipelines, inference may be invoked multiple times --- before, during, or after retrieval --- further compounding these costs, as we discuss in Section~\ref{sec:advanced_rag}.

\subsection{Vector Databases and Indexing}
\label{sec:vector_db}

\textbf{Building the Index.} Before retrieval can occur, all chunks of the document corpus are encoded into embedding vectors and stored in a vector database. Index creation is handled by the FlashRAG's preprocessing pipeline; however, the configuration choices made during this step have direct consequences for retrieval performance and memory footprint. The choice of embedding model determines vector dimensionality and encoding quality; the chunk size affects how much context is retrieved per passage; and the index type and numerical precision determine the trade-off between memory occupancy, search speed, and recall. In this work, we use the Flat index for our experiments while also examining the performance of retrieval-oriented indexing methods.

\textbf{Index Type:}  The choice of index structure determines how this search is performed:

\begin{itemize}
    \item \textit{Flat}: stores all embeddings explicitly and performs an exact brute-force search, guaranteeing full recall at the cost of linear scan time and full memory occupancy.
    \item \textit{IVF (Inverted File Index)}: partitions the embedding space into clusters and restricts search to the nearest clusters, reducing query time at the cost of some recall.
    \item \textit{SQ (Scalar Quantization)}: compresses embeddings by quantizing each dimension to a lower-precision scalar (e.g., FP32 to INT8), reducing memory footprint with minimal loss in recall compared to product quantization.
    \item \textit{IVF-SQ}: combines IVF clustering with scalar quantization, offering gains in both search speed and memory efficiency.
\end{itemize}

Approximate index structures may occasionally retrieve suboptimal chunks, which can degrade generation quality downstream.

\subsection{Advanced RAG Pipelines}
\label{sec:advanced_rag}

Modern RAG systems have evolved beyond the classical retrieve-then-generate paradigm to incorporate additional intermediate components. These can be broadly categorized by where they intervene in the pipeline: pre-retrieval methods, such as query rewriting~\cite{ma2023query} and web-search refine the input before retrieval, while post-retrieval methods, such as document re-ranking~\cite{nogueira2019passage} and prompt compression~\cite{jiang2023llmlingua, pan2024llmlingua, li2023compressing}, refine the retrieved context before generation. These methods rely on further LLM inference or encoding steps, increasing overall latency and memory footprint. However, the additional overhead can be offset by improvements in accuracy or reductions in downstream compute --- prompt compression, for instance, trims the input context to the generator, directly reducing generation cost.

Prompt compression, in particular, represents an important performance optimization opportunity in RAG. Hard prompt compression methods apply an auxiliary model to directly remove tokens from the retrieved context, keeping only tokens deemed important or query-relevant in order to reduce noise and accelerate downstream inference~\cite{jiang2023llmlingua, pan2024llmlingua}. However, the impact of compression is highly variable depending on the choice of compression method, generator model size, number of retrieved tokens, and compression rate --- making it a non-trivial configuration decision that requires systematic analysis and benchmarking to navigate.

RAG methods have also evolved in how and when retrieval occurs. Iterative RAG performs multiple passes through the retrieve-then-generate process to answer complex multi-hop questions, calling the generator multiple times with specific sub-goals on each iteration until a final answer is produced. Conditional RAG uses a separate model to evaluate question complexity and determine whether retrieval is necessary at all, skipping it entirely for simpler queries. In both cases, the number of inference calls becomes variable and workload-dependent, making the computational bottleneck difficult to characterize without benchmarking on representative data.
\section{Benchmark Implementation and Measurement Methodology}

\begin{figure}[ht]
    \centering
    \includegraphics[
        width=1\columnwidth,
        trim=1 5cm 0 1cm,
        clip
    ]{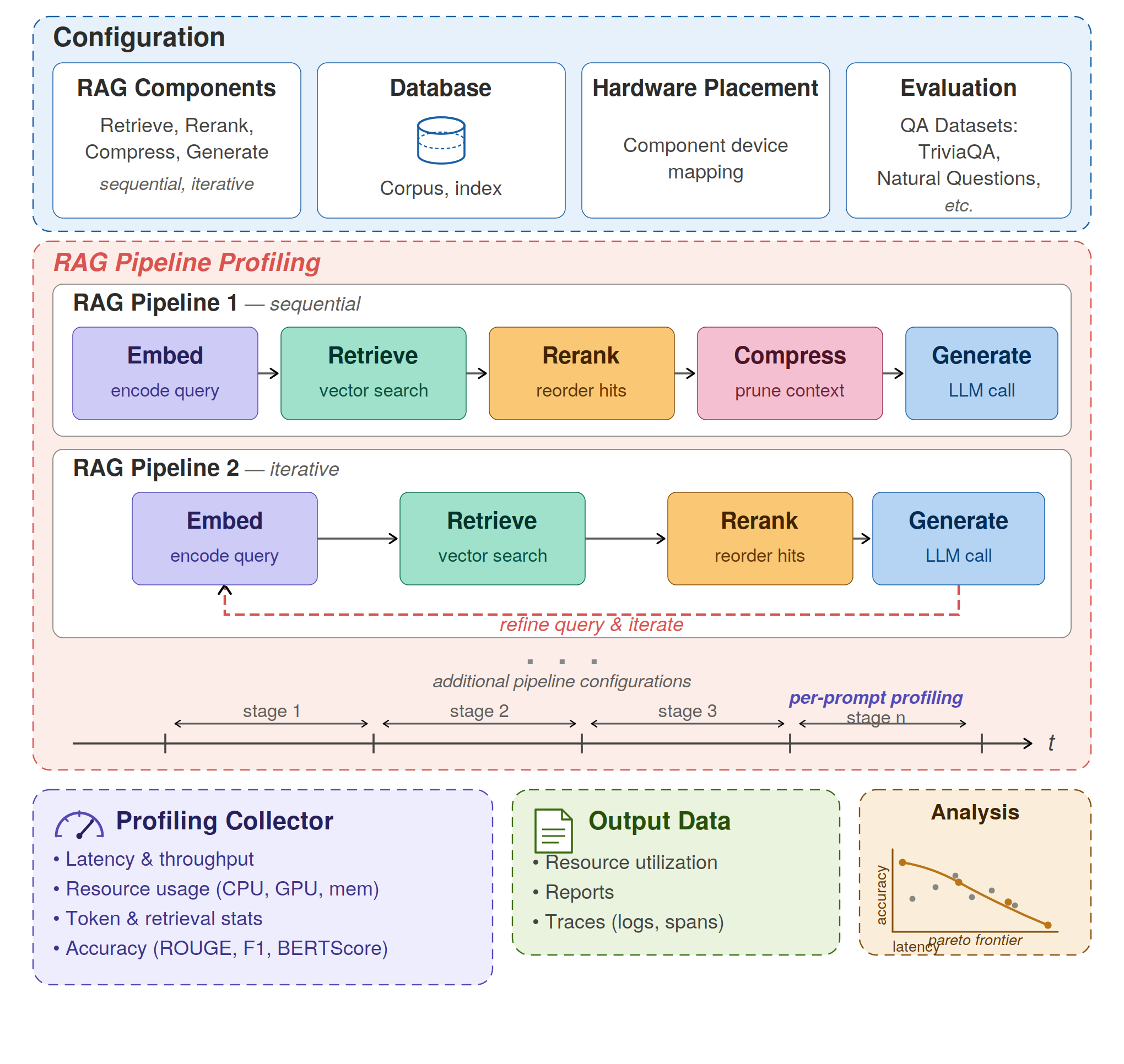}
    \caption{RAGMark overview. The framework takes configuration inputs (RAG components, database, hardware placement, evaluation datasets), executes one of two pipelines (sequential or iterative) with per-stage profiling, and produces resource, accuracy, and trace outputs for analysis.}
    \label{fig:RAGMark_overview}
\end{figure}

\subsection{Design Principles}

Our benchmarking framework follows five principles. First, \textit{representativeness}: it captures all stages of the RAG pipeline (embedding, retrieval, compression, and generation) on a per-query basis rather than treating RAG as a monolithic workload. Second, \textit{modularity}: the framework supports plug-and-play integration of retrievers, vector stores, compression methods, generator models, and datasets with minimal code changes. To maximize compatibility across components, it relies on standard Hugging Face interfaces~\cite{wolf-etal-2020-transformers} instead of specialized serving backends such as vLLM~\cite{kwon2023efficient}, since the goal is to characterize end-to-end RAG behavior rather than optimize serving throughput. Nevertheless, support for serving backends, such as vLLM, is under development to broaden the range of RAG deployment configurations. Third, \textit{joint evaluation}: the framework measures both system performance and model accuracy, enabling analysis of the trade-off between computational cost and answer quality often absent in prior work~\cite{jiang2025rago, jiang2024piperag}. Fourth, \textit{efficiency}: query sampling strategies allow meaningful characterization without evaluating every configuration on the full dataset. Finally, \textit{reproducibility}: all experiments are deterministic and configuration-driven, ensuring results can be reliably reproduced and extended.
A high level overview of our framework is illustrated in Figure~\ref{fig:RAGMark_overview}.

\subsection{Framework Implementation}

The framework is organized into a set of core classes, each representing a distinct component or implementation within the RAG pipeline, as summarized in Table~\ref{tab:classes}. 

\begin{table}[h]
\centering
\caption{Core framework classes and their responsibilities.}
\resizebox{\columnwidth}{!}{%
\begin{tabular}{ll}
\toprule
\textbf{Class} & \textbf{Responsibility} \\
\midrule

\multicolumn{2}{l}{\textit{RAG Stages}} \\
\midrule

\texttt{EmbeddingModel}
& Encodes queries via mean pooling over final hidden states \\
\multicolumn{2}{l}{\quad\small\textit{Key Parameters: model\_name, device, quantization}} \\[2pt]

\texttt{Retriever}
& Manages FAISS index loading, retrieval, GPU transfer, and sharding \\
\multicolumn{2}{l}{\quad\small\textit{Key Parameters: index\_path, top\_k, index\_type, gpu\_ids}} \\[2pt]

\texttt{Reranker}
& Re-scores retrieved passages using a cross-encoder \\
\multicolumn{2}{l}{\quad\small\textit{Key Parameters: model\_name, rerank\_k, device}} \\[2pt]

\texttt{Compressor}
& Unified interface over prompt compression backends \\
\multicolumn{2}{l}{\quad\small\textit{Key Parameters: method, compression\_rate, device}} \\[2pt]

\texttt{Generator}
& Causal LLM inference with instrumented latency measurement \\
\multicolumn{2}{l}{\quad\small\textit{Key Parameters: model\_path, max\_new\_tokens, quantization, device}} \\

\midrule

\multicolumn{2}{l}{\textit{Profiling}} \\
\midrule

\texttt{RAGTiming}
& Records per-stage latency and resource metrics for each query \\[2pt]

\texttt{PowerPoller}
& Samples GPU and CPU power draw at fixed intervals \\
\multicolumn{2}{l}{\quad\small\textit{Key Parameters: gpu\_ids, interval}} \\[2pt]

\texttt{TracePoller}
& Collects GPU utilization and memory traces \\
\multicolumn{2}{l}{\quad\small\textit{Key Parameters: gpu\_ids, interval}} \\

\midrule

\multicolumn{2}{l}{\textit{Pipeline}} \\
\midrule

\texttt{StandardPipeline}
& Single-pass retrieval with optional reranking and compression \\

\texttt{IterativePipeline}
& Multi-round retrieval with interleaved generation \\

\multicolumn{2}{l}{\quad\small\textit{Execution Modes: single-query, batch}} \\[2pt]

\multicolumn{2}{l}{\quad\small\textit{Key Parameters: embedder, retriever, reranker, compressor, generator}} \\

\bottomrule
\end{tabular}
}
\label{tab:classes}
\end{table}

The \texttt{Generator} class instruments generation with a custom
\texttt{LogitsProcessor} to measure time-to-first-token (TTFT) and per-token decode latency at the query level. The \texttt{Compressor} and \texttt{Reranker} classes provide unified interfaces for compression and reranking models: compression methods are normalized to a common compression-ratio parameter~$r$, while reranking methods are normalized to a rerank-keep parameter controlling the number of retained documents. The framework currently includes two pipeline subclasses, \texttt{StandardPipeline} and \texttt{IterativePipeline}.


\textbf{Sweep and Monitoring.} The framework is driven by \texttt{sweep.py}, which loads configuration definitions spanning database indexes, model choices, retrieval depth, compression methods, evaluation sizes, and component device placement before invoking the main evaluation driver, \texttt{main.py}. Within the evaluation pipeline, per-stage latency is recorded using \texttt{RAGTiming}, GPU and CPU power are sampled by \texttt{PowerPoller}, and hardware utilization traces are collected through \texttt{TracePoller}.

\textbf{Run Modes.} The framework supports two execution modes. In \textit{single-query} mode, queries are processed individually to enable fine-grained per-query latency analysis. In \textit{batch} mode, queries are
processed together across all stages that support batching, enabling throughput-oriented characterization.

\textbf{Hardware Deployment.} Each pipeline component (the retriever, database index, reranker, compressor, and generator) can be independently assigned to CPU or GPU resources via the configuration file. Further, the database index and generation model can be distributed across multiple devices. This enables systematic evaluation of heterogeneous deployment configurations.

\textbf{Integration with FlashRAG.} FlashRAG~\cite{jin2025flashrag} is an established RAG benchmark framework providing pipeline implementations, datasets, and evaluation metrics. We rely on FlashRAG for dataset download and preprocessing, FAISS~\cite{faiss_github} index creation, and QA evaluation infrastructure. However, FlashRAG is designed for accuracy evaluation rather than performance (i.e., speed) evaluation --- it processes all queries in bulk (retrieve-all, then embed-all, then generate-all), making per-query profiling impractical. Our benchmarking infrastructure and performance measurement pipeline are therefore implemented completely independently.

The major differences between RAGMark and existing infrastructure are that RAGMark provides the per-query, per-stage measurement pipeline, including stage-interleaved execution, fine-grained profiling, NVML-based energy integration and memory attribution, unified component interfaces, configuration-driven sweep orchestration, and unified per-query data analysis scripts. We leverage existing Hugging Face models, FAISS indexing, and corpus preprocessing infrastructure.

\textbf{Extensibility.}
Adding a generator, embedding model, or reranker to RAGMark only requires
specifying the Hugging Face model identifier and device in the
configuration file, while a FAISS database requires the index and
corpus locations. Adding a new component type (e.g., a query rewriter)
only involves implementing the corresponding component class and registering
it in the pipeline, with class templates provided by the framework.


\subsection{Measurement Methodology}
\label{sec:measurement}

Table~\ref{tab:metrics} summarizes the key metrics collected by the framework on a per-query basis across the categories that we describe next.

\subsubsection{Latency and Throughput}

The framework records wall-clock timestamps at stage boundaries to measure per-stage latency for embeddings, retrieval, compression, and LLM generation. Generation latency is further decomposed into time-to-first-token (TTFT) and decode time, capturing the distinct computational characteristics of the prefill and decode stages, respectively. End-to-end latency is reported as the sum of the measured latencies across all pipeline stages for each query. Latency and throughput can be collected on either a batch or single-query basis. 

\subsubsection{Memory and Resource Utilization}

Memory utilization is captured at two levels. At the device level, a background monitoring thread tracks true GPU memory consumption via the NVIDIA Management Library (NVML)~\cite{nvidia_nvml}, capturing all allocations, including FAISS indices alongside PyTorch tensors. At the framework level, PyTorch memory APIs track allocations managed by PyTorch directly~\cite{paszke2019pytorch}. Both peak and allocated GPU memory are logged per pipeline stage, enabling analysis of how memory footprint scales with configuration choices such as retrieval depth, embedding dimensionality, and model size. Host-side CPU utilization and process RSS memory are captured via \texttt{psutil}~\cite{psutil}.

\subsubsection{Energy Consumption}

GPU power and streaming multiprocessor (SM) occupancy are collected via a background monitoring thread using NVML, with power and SM occupancy sampled from GPU registers. Per-query energy is estimated by integrating sampled power readings over the query duration. This enables comparison of total energy consumption across pipeline configurations, model scales, and retrieval depths on a per-query basis.

\begin{table}[t]
\centering
\small
\caption{Examples of key per-qu{}ery metrics captured by the RAG benchmarking framework.}
\renewcommand{\arraystretch}{1.1}
\resizebox{\columnwidth}{!}{%
\begin{tabular}{lp{8.3cm}}
\toprule
\textbf{Category} & \textbf{Description and Example Measurements} \\
\midrule

Latency
& Per-stage and end-to-end query latency (embedding, retrieval, reranking, compression, generation TTFT, generation decode time, end-to-end latency) \\

Throughput
& Output generation throughput and response length (token throughput, generated tokens) \\

Context
& Input context size throughout the retrieval pipeline (tokens before/after retrieval, reranking, and compression) \\

GPU Memory
& PyTorch-managed and NVML-reported GPU memory usage (allocated memory, peak memory, NVML memory) \\

GPU Utilization \& Energy
& GPU activity and energy behavior during execution (power draw, energy consumption, SM utilization, memory bandwidth utilization) \\

CPU Metrics
& Host-side compute and memory footprint (CPU utilization, RSS memory) \\

Data Movement
& Host-device transfer activity (PCIe receive/transmit throughput) \\

\bottomrule
\end{tabular}
}

\label{tab:metrics}
\end{table}

\subsubsection{Accuracy Metrics}

For each configuration, the framework evaluates generated answers against ground-truth references using Exact Match (EM), which measures whether the generated query exactly matches the reference answer. We also include ROUGE, F1, and BERTScore, which are lightweight evaluation metrics that capture partial lexical overlap and semantic similarity between generated and reference text. These metrics are deterministic reference-based evaluation methods and do not require additional LLM inference during evaluation.

\subsection{Configuration Space}

A central goal of RAGMark is to enable systematic exploration of the RAG design space. Table~\ref{tab:config_space} summarizes key dimensions that can be directly specified as configuration parameters, though the framework is designed to extend beyond this. For example, integrating new generator models, rerankers, or embedding models typically requires only specifying the corresponding Hugging Face model identifier. Furthermore, parameters such as retrieval depth, re-ranking depth, compression rate, and quantization settings can be specified in the configuration.

\begin{table}[t]
\caption{RAG Benchmark Configuration Space}
\centering
\small
\renewcommand{\arraystretch}{1.2}
\begin{tabular}{lp{4.5cm}}
\toprule
\textbf{Axis} & \textbf{Values} \\
\midrule
Datasets & Wikipedia 2018, Wikipedia 2021, MS MARCO \\
Embedding Models & e5-small-v2, e5-base-v2, e5-large-v2 \\
Index Types & Flat (exact), IVF, IVF-SQ \\
Retrieval Depth & top-$k \in \{0, 1, 3, 5, 10\}$ where top-$k=0$ is LLM inference only \\
Reranking Model & ms-marco-MiniLM-L-6-v2 \\
Reranking Depth & top-$r \in \{3, 5\}$ retained after reranking \\
Generator Models & Llama-3.2-1B, Llama-3.2-3B, Llama-3.0-8B \\
Compression Methods & None, Selective-Context, LLMLingua, LLMLingua-2 \\
Compression Rates & $\{0.2, 0.4, 0.6, 0.8\}$ \\
Batch Size & $\{1, 2, 4, 8, 16\}$ \\
Hardware Placement & Per-component device assignment (retriever, vector index, compressor, generator) across CPU and GPU devices \\
Pipeline Strategy & Standard RAG, Iterative RAG \\
Iteration Rounds & $\{1, 2\}$  \\
Quantization and Precision & 4-bit, 8-bit, 16/32-bit \\
\bottomrule
\end{tabular}
\label{tab:config_space}
\end{table}

One key challenge of evaluating a large configuration
space is the required benchmark execution time, which scales with both the number
of configurations and the number of evaluated queries. Therefore, the
framework includes an evaluation-size parameter and sampling approach
to select a representative subset of queries, held consistent across
all configurations. In our experiments, we use a uniform random sample
of 400 QA pairs from each evaluation dataset, corresponding to
approximately 10\% mean coverage of the evaluation sets. Holding the
sample consistent across configurations enables direct comparisons,
while sampling, together with minimizing repeated model and database
initialization, keeps the configuration sweep practical.

\subsection{Experimental Setup}
\label{sec:datasets}

\subsubsection{Configuration Approach}

We divide our characterization study into five primary cases, as outlined in Table~\ref{tab:pipeline_configs}. The evaluated configuration space is summarized in Table~\ref{tab:config_space}. To isolate the impact of different RAG pipeline configurations under a consistent deployment environment, the retriever, generator, reranker, and compressor reside on GPU0, while the database is sharded across both GPU0 and GPU1. All experiments include a warm-up phase of five queries prior to measurement to ensure steady-state execution, and reported results are averaged across multiple runs for stability.

\subsubsection{Retrieval Corpus}

For the retrieval corpus, we use the compressed English Wikipedia XML dump from 2018 (approximately 14 GB, excluding user and talk pages). The corpus is pre-processed using FlashRAG~\cite{jin2025flashrag} scripts, using a soft limit of 512 words per chunk, resulting in approximately 9.2 million passages, averaging 215.5 words in length. Each passage is encoded using either e5-small-v2, e5-base-v2, or e5-large-v2, producing FAISS Flat indices of 13.9 GB, 28.2 GB, and 37.6 GB, respectively. 

\subsubsection{QA Evaluation Datasets}

We evaluate RAG accuracy across five open-domain QA benchmarks: Natural Questions (NQ)~\cite{kwiatkowski-etal-2019-natural}, TriviaQA~\cite{joshi-etal-2017-triviaqa}, SQuAD~\cite{rajpurkar-etal-2016-squad}, WebQuestions~\cite{berant-etal-2013-semantic}, and PopQA~\cite{mallen-etal-2023-when}. Additionally, we include HotpotQA~\cite{yang2018hotpotqa}, a multi-hop question benchmark that, in addition to single-hop queries, asks more complex questions, providing a comprehensive basis for evaluating how RAG configuration choices affect generation quality.

\section{Results}

\subsection{Platform Setup}
\label{sec:platform}

Our experiments are conducted on a multi-GPU node equipped with two NVIDIA A100-SXM4-40GB GPUs (40 GB VRAM each), running NVIDIA driver 545.23.08 and CUDA 12.3. The A100 provides 312 TFLOPS FP16 via third-generation tensor cores. The host system features a 48-core Intel Xeon Gold 6240R CPU running at 2.40 GHz.

\begin{table}[t]
\caption{RAG Case Studies}
\label{tab:cases}

\centering
\renewcommand{\arraystretch}{1.2}
\resizebox{\columnwidth}{!}{%
\begin{tabular}{lp{5cm}}
\toprule
\textbf{Case} & \textbf{Pipeline Components} \\
\midrule
Case 1: Basic & Embedding, FAISS retrieval, generation \\
Case 2: Reranking & Embedding, FAISS retrieval, cross-encoder reranking, generation \\
Case 3: Compression & Embedding, FAISS retrieval, prompt compression, generation \\
Case 4: Combined & Embedding, FAISS retrieval, cross-encoder reranking, prompt compression, generation \\
Case 5: Iterative RAG & Iterative embedding, FAISS retrieval, and generation over multiple rounds \\
\bottomrule
\end{tabular}
}
\label{tab:pipeline_configs}
\end{table}

\subsection{Case 1: Naive Pipeline Analysis}

\noindent\textbf{Latency:} 
We first analyze the execution time breakdown across RAG pipeline stages in the basic retrieval configuration. LLM TTFT and decode dominate end-to-end latency across all models and retrieval depths, while retrieval and embedding contribute under 35~ms, even at \texttt{topk=10}. FAISS latency grows from $\sim$7~ms at \texttt{topk=1} to $\sim$27~ms at \texttt{topk=5}, and then increases marginally afterward, suggesting that the main computational cost lies in the search operation itself rather than in retrieving additional nearest neighbors. Embedding stays near 7~ms throughout.

As expected, TTFT scales with retrieval depth (token count) for larger models. \texttt{L3-8B} grows from 64~ms at \texttt{topk=0} (standard LLM inference), to 185~ms at \texttt{topk=5} and 322~ms at \texttt{topk=10}, with the increment from \texttt{topk=5} to \texttt{10} (+137~ms) exceeding the increment from \texttt{topk=0} to \texttt{5}, despite covering the same number of additional documents. This reflects attention's $O(n^2)$ scaling: at short contexts, prefill is dominated by linear MLP work, but the quadratic attention term becomes dominant as context grows. \texttt{L3.2-3B} shows the same TTFT (53~ms, 103~ms, 170~ms) pattern, while \texttt{L3.2-1B} remains linear (32~ms, 46~ms, 70~ms) because its smaller hidden dimension keeps attention cheap. Decode latency growth is model-dependent: the 8B model's decode more than doubles from 84~ms to 204~ms across \texttt{topk=0} to \texttt{10}, while the 1B latency remains nearly flat (42~ms to 69~ms). 




\begin{table*}[t]
\caption{RAG resource usage across four different pipeline configurations.}
\centering
\resizebox{\textwidth}{!}{%
\begin{tabular}{lllrrrr|rrrr|rrrr|rrrr}
\toprule
 &  &  & \multicolumn{4}{c}{Case 1: Naive} & \multicolumn{4}{c}{Case 2: Rerank} & \multicolumn{4}{c}{Case 3: Compression} & \multicolumn{4}{c}{Case 4: Combined} \\
 Model & top-K  & GPU ID & J & GB & SM\% & mem\% & J & GB & SM\% & mem\% & J & GB & SM\% & mem\% & J & GB & SM\% & mem\% \\

\midrule
\multirow[t]{6}{*}{L3.2-1B} & \multirow[t]{2}{*}{0} & GPU0 & 8.40 & 3.91 & 32.14 & 10.82 & -- & -- & -- & -- & -- & -- & -- & -- & -- & -- & -- & -- \\
 &  & GPU1 & 6.32 & 1.02 & 0.00 & 0.00 & -- & -- & -- & -- & -- & -- & -- & -- & -- & -- & -- & -- \\
\midrule
\cline{2-19}
 & \multirow[t]{2}{*}{5} & GPU0 & 19.61 & 11.63 & 43.79 & 14.59 & 17.64 & 11.79 & 43.41 & 13.55 & 16.52 & 12.97 & 41.61 & 10.58 & 15.58 & 13.02 & 41.69 & 10.24 \\
 &  & GPU1 & 9.05 & 7.50 & 13.88 & 2.27 & 8.52 & 7.50 & 13.31 & 2.18 & 7.51 & 7.50 & 11.75 & 1.86 & 7.20 & 7.50 & 11.92 & 1.91 \\
\midrule
\cline{2-19}
 & \multirow[t]{2}{*}{10} & GPU0 & 25.41 & 12.09 & 45.87 & 17.12 & 18.55 & 11.89 & 43.44 & 13.22 & 21.06 & 12.64 & 41.98 & 10.68 & 16.23 & 13.09 & 41.97 & 10.12 \\
 &  & GPU1 & 11.19 & 7.50 & 11.16 & 1.77 & 8.68 & 7.50 & 12.92 & 2.08 & 7.90 & 7.50 & 9.26 & 1.43 & 7.27 & 7.50 & 11.34 & 1.78 \\
\midrule
\cline{1-19} \cline{2-19}
\midrule
\multirow[t]{6}{*}{L3.2-3B} & \multirow[t]{2}{*}{0} & GPU0 & 22.58 & 7.65 & 39.50 & 18.79 & -- & -- & -- & -- & -- & -- & -- & -- & -- & -- & -- & -- \\
 &  & GPU1 & 11.35 & 1.02 & 0.00 & 0.00 & -- & -- & -- & -- & -- & -- & -- & -- & -- & -- & -- & -- \\
\midrule
\cline{2-19}
 & \multirow[t]{2}{*}{5} & GPU0 & 49.28 & 17.53 & 54.25 & 22.95 & 42.75 & 15.65 & 53.14 & 21.48 & 35.07 & 17.31 & 48.41 & 16.72 & 32.34 & 16.70 & 47.62 & 16.15 \\
 &  & GPU1 & 17.55 & 7.94 & 8.78 & 1.40 & 15.85 & 7.50 & 9.20 & 1.45 & 13.35 & 7.64 & 8.84 & 1.38 & 12.75 & 7.50 & 9.09 & 1.43 \\
\midrule
\cline{2-19}
 & \multirow[t]{2}{*}{10} & GPU0 & 75.50 & 18.14 & 58.12 & 25.59 & 43.65 & 15.69 & 53.17 & 21.12 & 49.54 & 17.09 & 48.94 & 16.21 & 32.84 & 16.89 & 47.88 & 15.86 \\
 &  & GPU1 & 22.95 & 7.94 & 6.81 & 1.05 & 16.06 & 7.50 & 8.83 & 1.37 & 15.53 & 7.64 & 7.85 & 1.21 & 12.62 & 7.49 & 8.74 & 1.34 \\
\midrule
\cline{1-19} \cline{2-19}
\midrule
\multirow[t]{6}{*}{L3-8B} & \multirow[t]{2}{*}{0} & GPU0 & 29.96 & 17.61 & 59.23 & 37.53 & -- & -- & -- & -- & -- & -- & -- & -- & -- & -- & -- & -- \\
 &  & GPU1 & 10.94 & 1.02 & 0.00 & 0.00 & -- & -- & -- & -- & -- & -- & -- & -- & -- & -- & -- & -- \\
\midrule
\cline{2-19}
 & \multirow[t]{2}{*}{5} & GPU0 & 87.08 & 24.65 & 70.25 & 36.09 & 74.35 & 24.80 & 68.46 & 34.69 & 56.54 & 25.62 & 60.61 & 28.19 & 50.72 & 25.77 & 59.48 & 27.82 \\
 &  & GPU1 & 22.35 & 7.50 & 6.81 & 1.08 & 19.87 & 7.50 & 7.15 & 1.11 & 15.92 & 7.50 & 7.55 & 1.17 & 15.11 & 7.50 & 7.87 & 1.23 \\
\midrule
\cline{2-19}
 & \multirow[t]{2}{*}{10} & GPU0 & 157.52 & 25.49 & 72.42 & 35.49 & 75.51 & 24.89 & 67.61 & 33.88 & 85.65 & 25.53 & 59.54 & 25.28 & 52.92 & 25.92 & 59.45 & 27.58 \\
 &  & GPU1 & 36.27 & 7.50 & 4.81 & 0.74 & 19.98 & 7.50 & 7.07 & 1.09 & 21.03 & 7.50 & 6.61 & 1.02 & 15.18 & 7.50 & 7.74 & 1.19 \\
\midrule
\cline{1-19} \cline{2-19}
\bottomrule
\end{tabular}
}
\label{tab:combined_metrics}
\end{table*}

\begin{figure*}[t]
    \centering
    \includegraphics[width=0.85\textwidth]{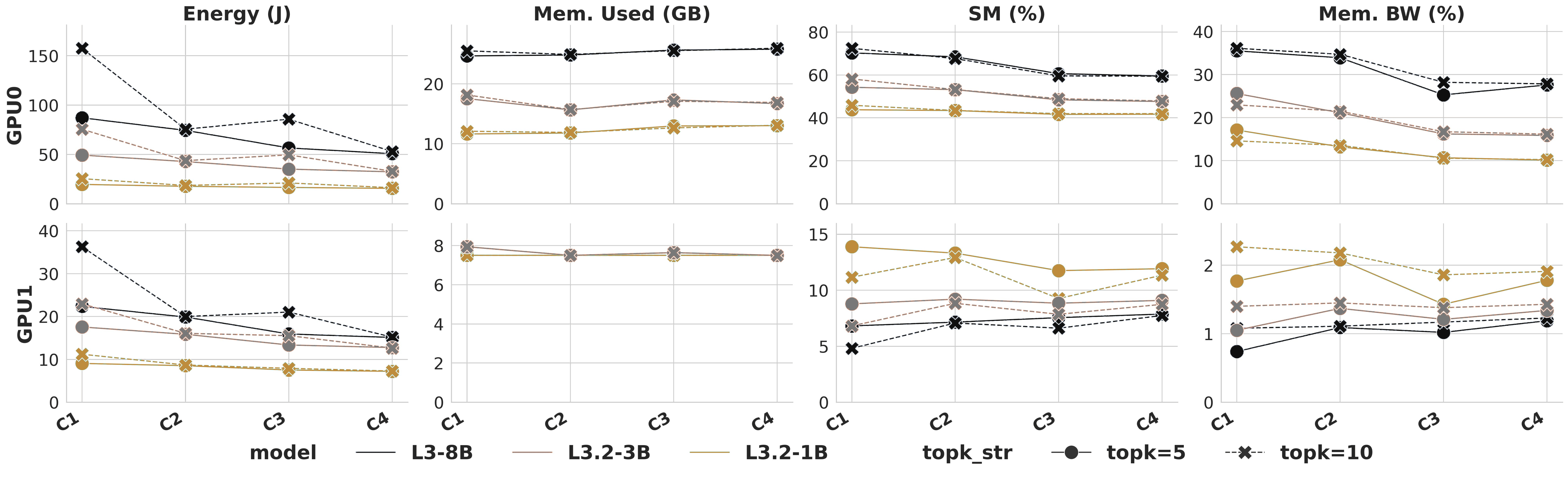}
    \caption{Resource utilization trends across four RAG pipeline configurations.}
    \label{fig:utilization_plot}
\end{figure*}

\noindent\textbf{Resource Utilization:} GPU0 handles the entire generation workload, while GPU1 stays under 5\% SM utilization across all configurations (see Figure~\ref{fig:utilization_plot}). For \texttt{L3-8B}, SM utilization saturates as context grows, rising from 59.2\% at \texttt{topk=0} to 70.3\% at \texttt{topk=5} and 72.4\% at \texttt{topk=10}. But energy continues to climb from 87~J to 158~J across the same range. Saturated compute occupancy, with continued energy growth, signals an increasing fraction of time spent in the memory-bandwidth-bound decode region, where SMs remain busy, though are bottlenecked on data movement.

\noindent\textbf{Memory:} The peak GPU0 footprint is dominated by model parameters. The \texttt{topk=0} no-rag baseline allocates 3.9, 7.7, and 17.6~GB for \texttt{L3.2-1B}, \texttt{L3.2-3B}, and \texttt{L3-8B}, respectively, as shown in Table~\ref{tab:combined_metrics}. Adding retrieval increases the footprint by roughly 8~GB across all model sizes, increasing GPU0's allocation for \texttt{topk=10} to 12.1, 18.1, and 25.5~GB for \texttt{L3.2-1B}, \texttt{L3.2-3B}, and \texttt{L3-8B}, respectively. The increase reflects hosting the partial FAISS index on GPU0 alongside the generator. Memory bandwidth utilization for \texttt{L3-8B} remains in the 35--37\% range across retrieval depths, consistent with decode being bandwidth-limited, regardless of context length. GPU1's footprint remains constant at 7.5~GB (the partial FAISS index), with bandwidth utilization under 2\%, confirming that retrieval uses less compute than the generator. 

\noindent\textbf{Accuracy:} ROUGE scores improve with retrieval depth and stabilize beyond \texttt{topk=3}, confirming that retrieval delivers its expected quality benefit across the configurations studied.

\textbf{Key Insight:} The naive RAG pipeline transitions between different architectural bottlenecks as retrieval depth grows. At small contexts, prefill is compute-bound and decode involves a reduction in the KV cache cost. At larger contexts, attention's quadratic scaling saturates tensor cores, while decode becomes bandwidth-bound. The result is disproportionate energy scaling on larger models (1.8$\times$ growth from \texttt{topk=5} to \texttt{topk=10} on \texttt{L3-8B}), despite SM utilization plateauing: the system spends more time waiting on memory per unit of compute progress.


\subsection{Case 2: Reranking Analysis}

We next evaluate the reranking configuration to characterize how filtering retrieved documents before generation impacts the workload. 

\begin{figure*}[t]
    \centering
    \includegraphics[width=0.85\textwidth]{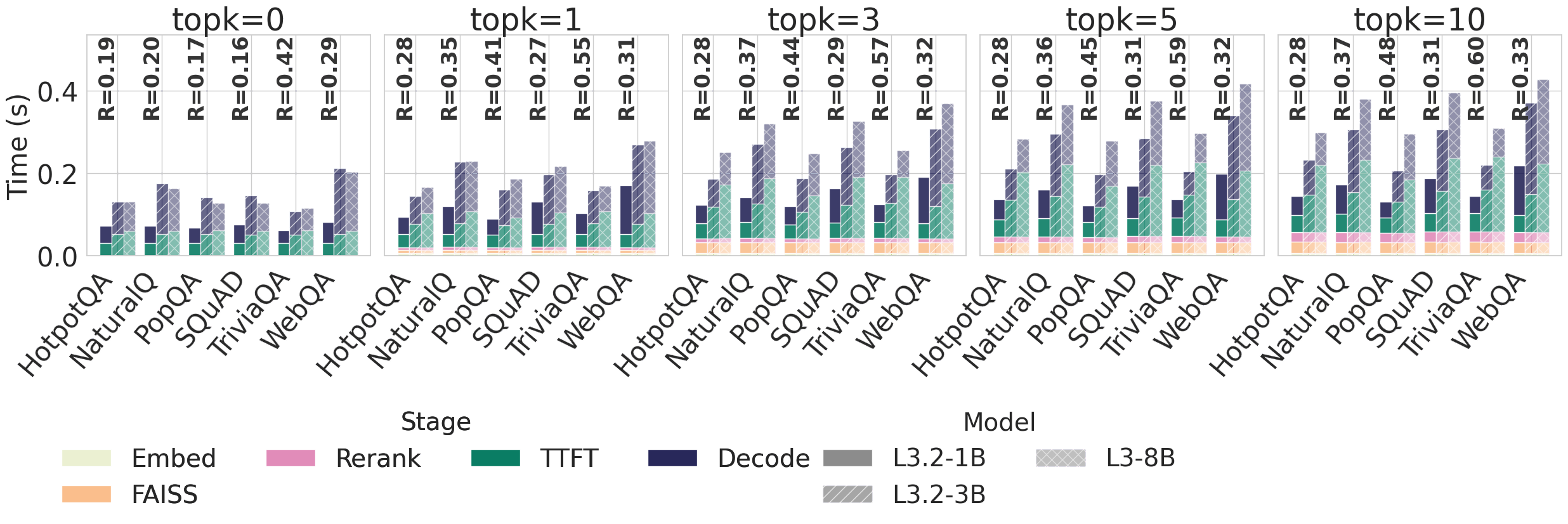}
    \caption{Latency breakdown and ROUGE scores for reranking (Case 2). }
    \label{fig:rouge_reranking_latency}
\end{figure*}

\noindent\textbf{Latency:} Latency results are summarized in Figure~\ref{fig:rouge_reranking_latency}, and show that reranking overhead scales linearly with retrieval depth, from $\sim$7~ms at \texttt{topk=1} to $\sim$24~ms at \texttt{topk=10}, reflecting the per-document cross-encoder scoring cost. By truncating the context passed to the generator, reranking moves prefill back toward the linear region: \texttt{L3-8B} TTFT drops from 322~ms to 165~ms at \texttt{topk=10}, a 49\% reduction that more than offsets the 24~ms reranking overhead. \texttt{L3.2-3B} sees a 45\% TTFT reduction (170~ms to 93~ms), while \texttt{L3.2-1B}, already in the linear region, sees a smaller absolute reduction (70~ms to 42~ms). Reranking benefits are greatest precisely where the naive pipeline suffers the most: large models at high retrieval depths. Decode follows the same logic, dropping from 204~ms to 128~ms on \texttt{L3-8B} at \texttt{topk=10}, as a smaller KV cache is traversed per generated token.

\noindent\textbf{Energy and SM Utilization:} Reranking impacts resource utilization significantly. For \texttt{L3-8B} at \texttt{topk=10}, GPU0 SM utilization decreases from 72.4\% to 67.6\%, and energy falls from 157.5~J to 75.5~J, a 52\% reduction which is far larger than the SM change, because energy integrates over a shorter generation period (see Figure~\ref{fig:utilization_plot}, C1 to C2). GPU1's SM utilization remains under 8\%, confirming that retrieval remains unaffected since the reranker is running on GPU1.

\noindent\textbf{Memory:} Reranking has some effect on the peak memory
footprint. At \texttt{topk=10} on \texttt{L3-8B}, GPU0 uses 25.5~GB
in the naive pipeline versus 24.9~GB with reranking. The reranker adds
approximately 200~MB of memory overhead, but the reduced generation
context decreases context-dependent memory usage, including KV
cache pressure, resulting in a small net reduction in peak memory. Memory
bandwidth utilization shows a small drop, from 35.5\% to 33.9\%,
reflecting faster KV cache traversals during decode rather than a
change in the bandwidth-bound nature of the workload. The footprint
remains largely dominated by model weights.

\noindent\textbf{Accuracy:} ROUGE scores differ by less than 0.02 from the naive pipeline across all configurations, confirming that the latency and energy gains do not sacrifice quality.

\noindent\textbf{Key Insight:} Reranking trades a modest increase in retrieval-side overhead (linear in topk, capped at $\sim$24~ms) for substantial generation-side savings (49\% TTFT reduction on L3-8B at topk=10). This is achieved by reducing the attention-quadratic prefill stage and shrinking the KV cache size during the decode stage. This optimization substantially reduces time and energy, while having only a small effect on peak GPU memory, which remains largely dominated by generator model weights.

%


\subsection{Case 3: Compression Analysis}

\begin{figure}[t]
    \centering
    \includegraphics[width=\columnwidth]{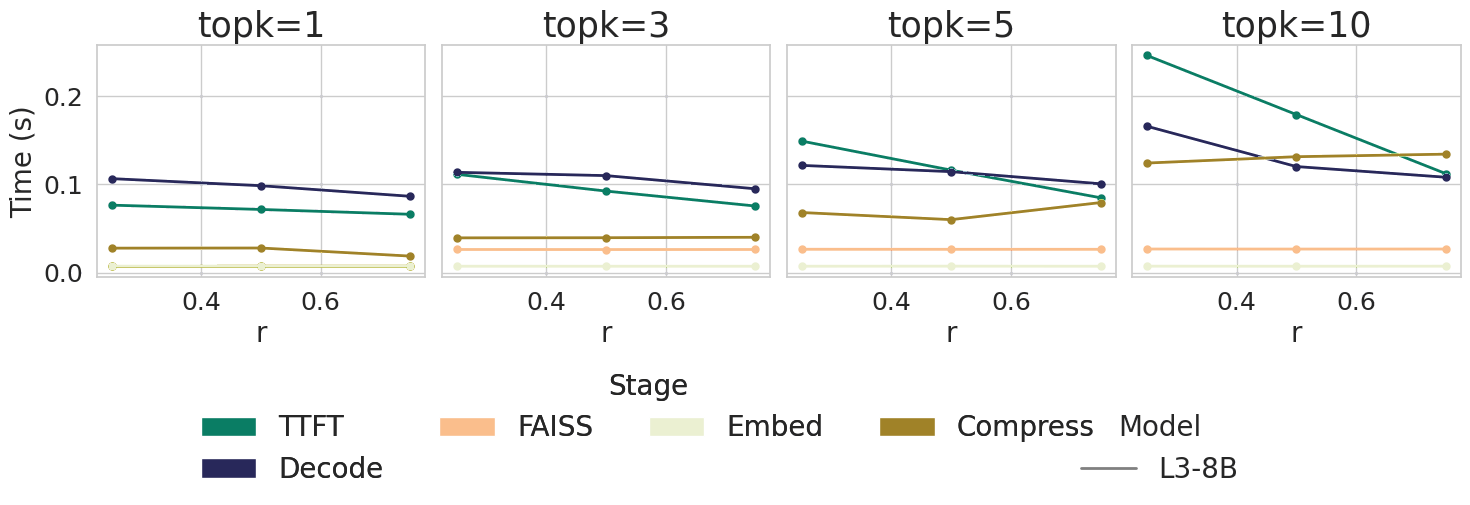}
    \caption{Impact of compression rate (r) on stage latencies using LLMLingua-2 (LLmng2) compression model.}
    \label{fig:compress_rates}
\end{figure}

\begin{figure}[t]
    \centering
    \includegraphics[width=\columnwidth]{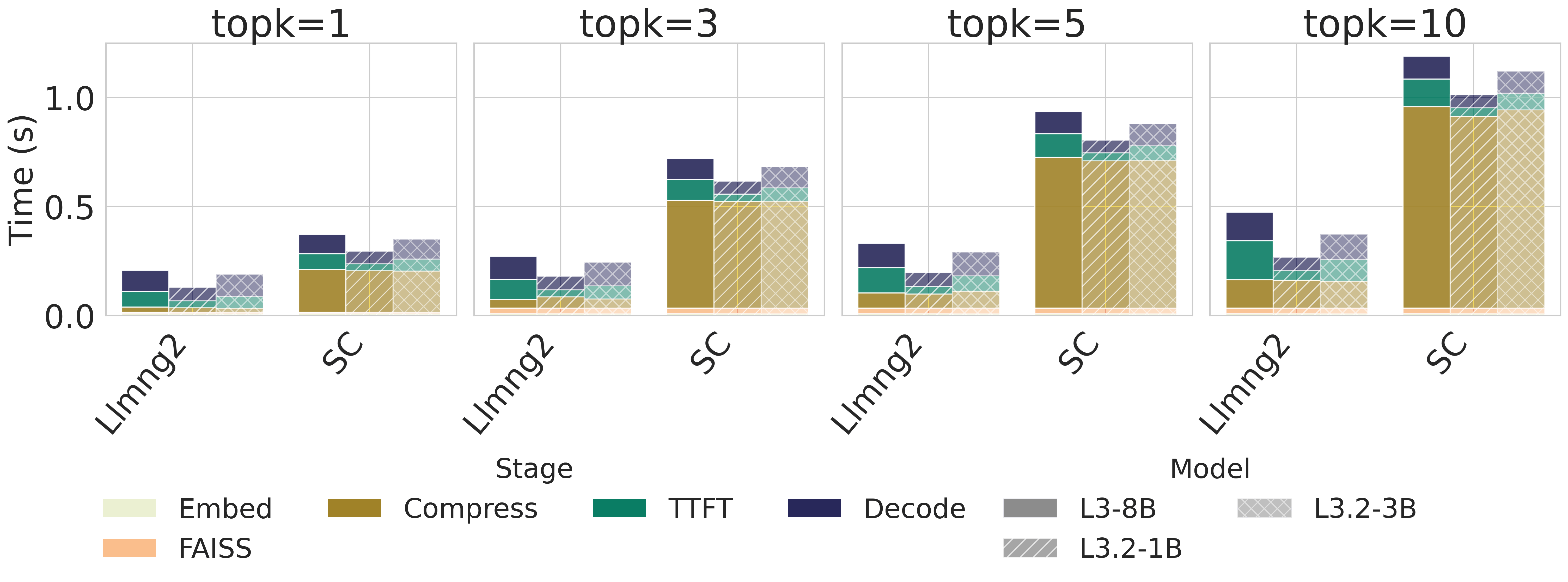}
    \caption{Compression latency overhead comparison between SC and LLMLingua-2.}
    \label{fig:output_compress_models}
\end{figure}

We evaluate the effect that adding an LLMLingua-2 (LLmng2) compressor has on latency and utilization. 
\noindent\textbf{Latency:} Compression overhead scales with retrieval depth, growing from $\sim$22~ms at \texttt{topk=1} to $\sim$58--64~ms at \texttt{topk=10} across models, on par with FAISS retrieval. The compressor runs as a transformer-based classifier on GPU0, sharing the device with the generator. The net system-level performance gain exceeds reranking's at high retrieval depths: for \texttt{L3-8B} at \texttt{topk=10}, TTFT drops from 322~ms to 108~ms, a 66\% reduction, which is larger than reranking's 49\%, and decode falls from 204~ms to 111~ms. Compression outperforms reranking here because it shrinks context both across and within documents: reranking discards whole passages, while compression additionally densifies the ones it keeps, compounding the prefill and KV cache reduction. Figure~\ref{fig:compress_rates} further breaks down latency by compression rate. As compression increases, TTFT shows the largest reduction, followed by decode latency, particularly for larger models. At the same time, higher compression rates increase the compute overhead of the compression step itself. The choice of compression method is also important. Figure~\ref{fig:output_compress_models} shows that Selective-Context (SC) is not beneficial in our pipeline because its latency overhead outweighs its gains. Therefore, all subsequent analysis uses LLMLingua-2.

\noindent\textbf{Resource Utilization:} Compression yields the larger energy optimization compared to reranking. For \texttt{L3-8B} at \texttt{topk=10}, GPU0's energy falls from 157.5~J to 85.7~J (a 46\% reduction, despite compression adding 58~ms of overhead to a shared GPU), and SM utilization drops from 72.4\% to 59.5\% (see Figure~\ref{fig:utilization_plot}, C1 to C3). The combination indicates compression moves the generator out of the saturated-compute region: prefill is small enough that tensor cores are no longer the bottleneck. The architectural caveat is that compression overhead is itself compute-bound and contends with the generator on GPU0; at smaller models sizes and retrieval depths, the compression cost can erase the generation savings.

\noindent\textbf{Memory:} Compression leaves the peak GPU0 footprint nearly unchanged at 25.5~GB on \texttt{L3-8B}, identical to the naive pipeline, because the compressor model itself occupies VRAM on GPU0 alongside the generator, offsetting the KV cache reduction. Memory bandwidth utilization, however, drops sharply, from 35.5\% to 25.3\% on \texttt{L3-8B} at \texttt{topk=10}, the largest bandwidth reduction of any single optimization (see Figure~\ref{fig:utilization_plot}). 

\noindent\textbf{Key Insight:} Similar to reranking, compression trades preprocessing overhead for generation-side savings by reducing the attention-quadratic prefill stage and shrinking the KV cache during decode, with larger gains (66\% TTFT, 46\% decode on L3-8B at topk=10) because it shrinks context both across and within passages. The compute cost depends heavily on the compressor choice (e.g., SC is much more expensive than LLMLingua2) and contends with the generator on the same GPU, so the trade only pays off when context is long enough to amortize it.
\subsection{Case 4: Composition Effects: Reranking + Compression}

We combine reranking and compression to evaluate whether their system benefits are complementary. The pipeline filters retrieved documents before compression, applying both context-reduction mechanisms in sequence.

\noindent\textbf{Latency:} The combined pipeline outperforms compression-only (C3), despite incurring additional reranking overhead. For \texttt{L3-8B} at \texttt{topk=10}, TTFT drops from 179~ms (compression-only) to 108~ms and decode from 131~ms to 111~ms, both well below the naive baseline (322~ms and 204~ms). The expected additive overhead does not overshadow the benefits: reranking shrinks the input to the compressor, and also cuts compression cost roughly in half (130~ms to 58~ms) and more than offsetting the cost of reranking (24~ms).

\noindent\textbf{Resource Utilization:} The combined pipeline achieves the lowest GPU0 energy. For \texttt{L3-8B} at \texttt{topk=10}, energy drops to 52.9~J, versus 75.5~J for reranking alone, 85.7~J for compression alone, and 157.5~J for the naive pipeline, a 66\% reduction from naive. The same pattern holds across model sizes (see Figure~\ref{fig:utilization_plot}, C4 column). Case 4 consistently utilizes less energy compared to Cases 2 and 3, as seen in Figure~\ref{fig:utilization_plot} with the gap widening at larger model sizes. Peak GPU0 footprint remains at 25.9~GB, dominated by model weights and the resident compressor and reranker.

\textbf{Key Insight.} Reranking and compression are not additive in cost, but compounding in benefit. Reranking shrinks the compressor's input, compression densifies what reranking retains, and the generator inherits a context reduced twice. The result is the largest energy reduction in our analysis (66\% on \texttt{L3-8B} at \texttt{topk=10}), achieved by stacking two stages that each look individually like added overhead.

\subsection{Case 5: Iterative Pipeline Analysis}

\begin{figure}[t]
    \centering
    \includegraphics[width=\columnwidth]{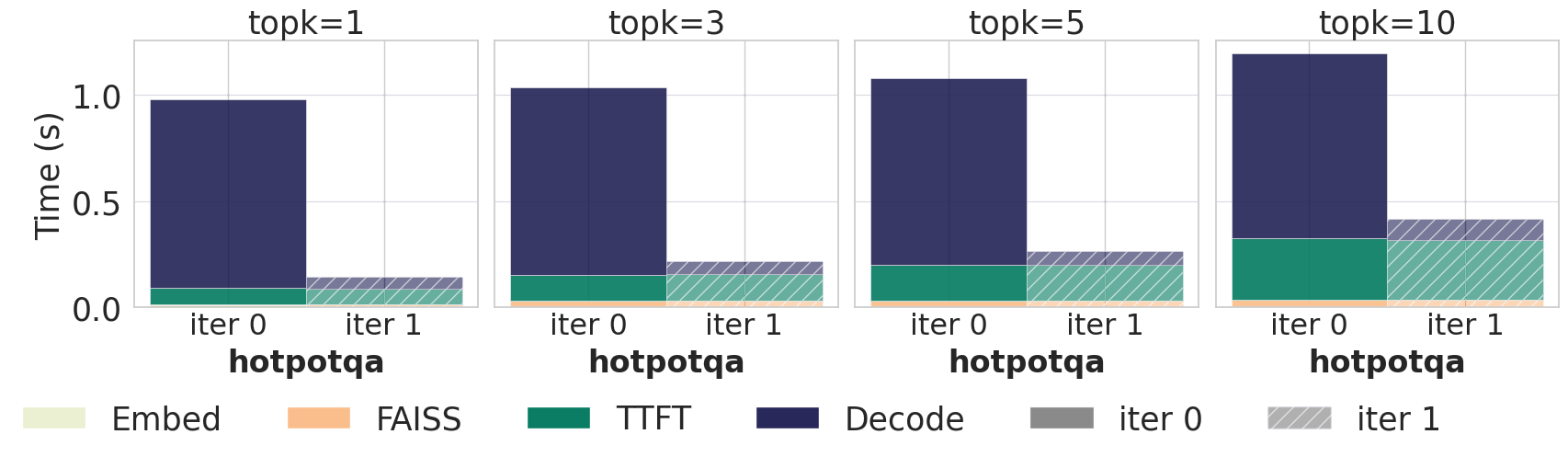}
    \caption{Iterative Pipeline Analysis.}
    \label{fig:iterative_latency}
\end{figure}

As an additional RAGMark use case, we change our processing paradigm, shifting to iterative RAG. Inspecting the latency trends across iterative generation passes in Figure~\ref{fig:iterative_latency}, we see that the pipeline incurs retrieval and embedding overhead twice, once for each retrieval-generation iteration. Despite this additional retrieval cost, decode latency is substantially larger during the first iteration (iter 0), while the second iteration (iter 1) exhibits much shorter decode times due to differences in prompt structure. Specifically, the first prompt instructs the model to generate an expanded answer, whereas the second prompt requests a short factual response. For example, in the L3-8B model at topk=10, the first-iteration decode latency reaches 867 ms, while second-iteration decode latency decreases substantially to 102 ms. In contrast, TTFT remains relatively the same between iterations, only changing from 291 ms to 281 ms. Similar trends are observed across the 1B and 3B models.

\textbf{Key Insight.} Iterative RAG fundamentally changes workload scaling behavior, amplifying both retrieval and generation costs, and introducing temporal dependencies that are absent in single-pass
pipelines.


\subsection{Index Methods}

\begin{figure}[t]
    \centering
    \includegraphics[width=1.0\linewidth, trim=0 0.0cm 0 0.0cm, clip]{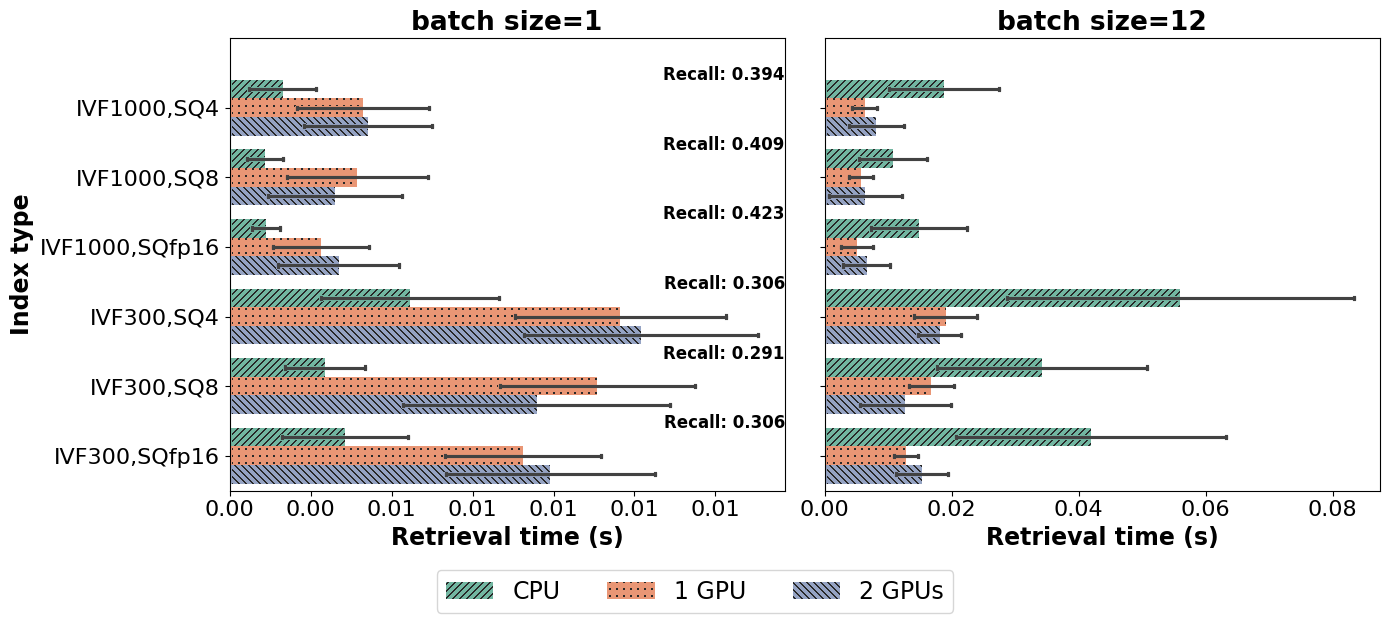}
    \caption{Index format impact on latency and accuracy across different device allocations.}
    \label{fig:RAG_index_study}
\end{figure}

Figure~\ref{fig:RAG_index_study} shows that retrieval latency depends strongly on hardware, indexing method, and batch size. For single-query prompts (batch=1), CPU-based IVF1000 indexes outperform GPU, with IVF1000,SQ8 reaching $\sim$1.1~ms versus $\sim$3.9~ms on a single GPU --- highlighting GPU overhead and transfer cost for lightweight queries. CPU latency remains lower than GPU latency across nearly all low-batch configurations, especially for quantized IVF1000 indexes.

GPU implementations scale better with batch size. At batch=12, CPU latency for IVF300,SQ4 increases to $\sim$56~ms, while the single-GPU equivalent remains near 34~ms, with similar trends across other IVF300 indexes. A second GPU does not always help due to synchronization overhead, but GPUs generally scale better for batched retrieval.

Memory footprint also matters: for the Wikipedia 2018 dataset with E5-small, compressed IVF indexes require only a few GBs versus the much larger flat index. CPUs remain attractive for large databases and deployments where inference already consumes most VRAM, while GPUs become preferable as retrieval concurrency and workload size increase.

\textbf{Key Insight.} Hardware placement can invert the workload profile: the input query, query batching and database size can affect decisions about how to allocate resources to the database index.


\subsection{Evaluating Trade-offs}

\begin{figure}[htbp]
    \centering
    \includegraphics[width=0.95\linewidth, trim=0 0.2cm 0 0.0cm, clip]{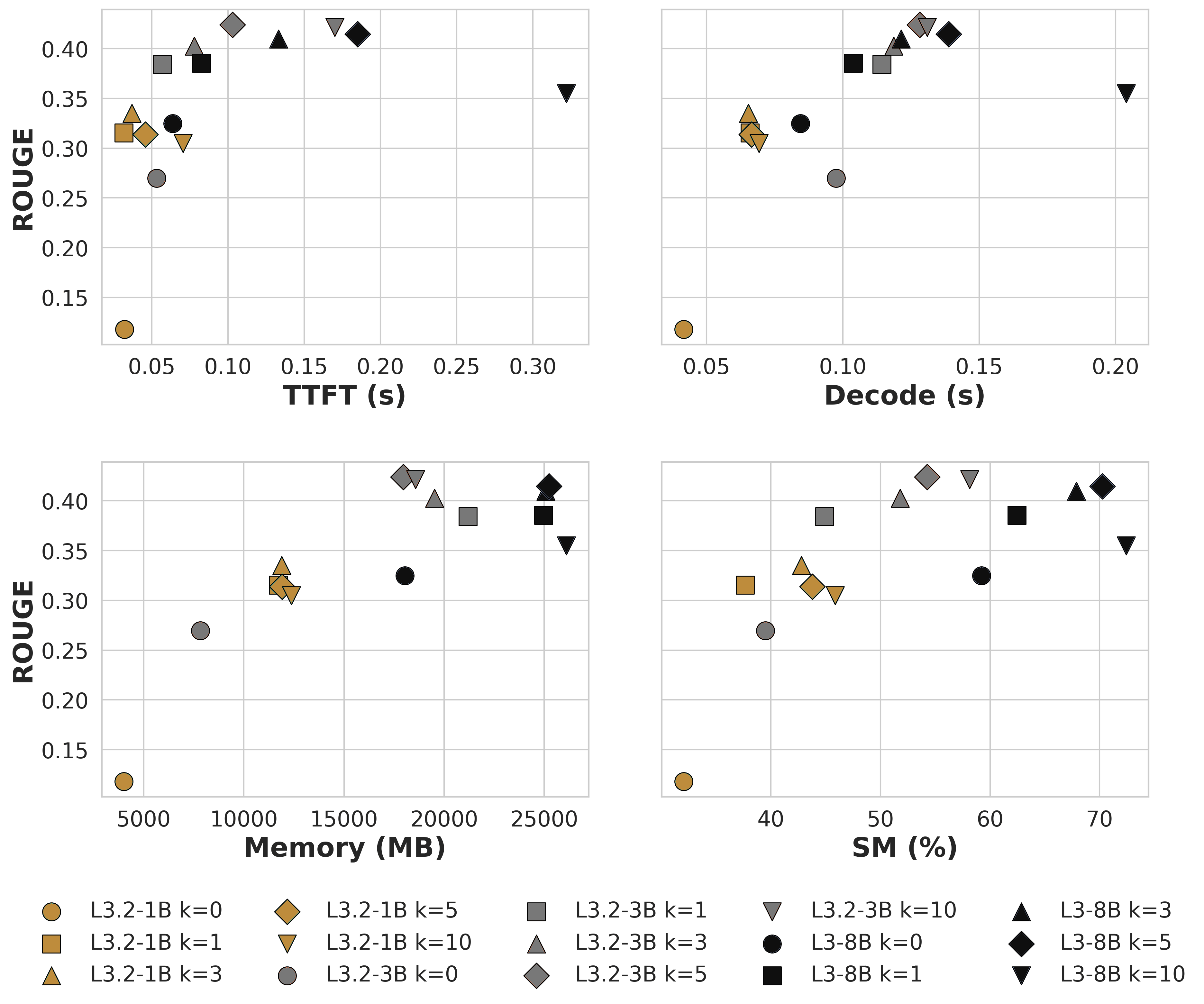}
    \caption{Trade-offs between system metrics and ROUGE scores for the naive RAG pipeline (Case 1).}
    \label{fig:optimization_viz}
\end{figure}

The data generated by our benchmark allows users to easily explore trade-offs between RAG accuracy metrics and resource utilization. Here we present a short motivating example that visualizes the trade-offs in Case 1: the naive pipeline. 
In Figure~\ref{fig:optimization_viz}, we can observe the trade-offs between the TTFT and ROUGE score. The 1B model with k=3 and k=5 achieve surprisingly competitive ROUGE score ($\sim$0.33), at a very low TTFT ($\sim$0.04s), which is the lowest RAG latency. With k=8, the 8B model achieves the highest ROUGE score ($\sim$0.41), but with $\sim$5-8x higher latency. Increasing k from 3 to 10 does not consistently improve ROUGE, but does increases TTFT noticeably, especially for larger models.

\section{Related Work}

RAG systems sit at the intersection of information retrieval and LLM inference. Prior work has largely focused on two directions: i) optimizing RAG execution and efficiency, and ii) workload characterization and benchmarking.

\noindent\textbf{RAG Optimization.}
Recent work has explored improving RAG efficiency through pipeline parallelism, caching, and resource-aware scheduling. PipeRAG~\cite{jiang2024piperag} and Teola~\cite{tan2024teola} decompose RAG workflows into finer grained stages and executes them in a pipelined fashion, reducing latency and increasing utilization. Other systems focus on caching and reuse across repeated queries~\cite{jin2024ragcache, yao2025cacheblend}, while resource-aware frameworks~\cite{jiang2025rago, jiang2023chameleon} study placement and orchestration strategies. These efforts focus on performance analysis and optimization rather than delivering reusable benchmarking infrastructure.

\noindent\textbf{Workload Characterization and Benchmarking.}
Despite the growing deployment of RAG systems, comprehensive hardware benchmarking for RAG remains limited. Existing analyses largely focus on isolated LLM inference~\cite{zhou2024survey, agrawal2024vidur, niu2025tokenpowerbench} without modeling the pre- and post-retrieval stages that distinguish RAG from standard generation workloads. FlashRAG~\cite{jin2025flashrag} and BERGEN~\cite{rau2024bergen} provide modular frameworks for evaluating RAG accuracy across pipelines and datasets, but do not focus on per-stage metrics. Shen et al.~\cite{shen2024towards} characterize TTFT and tail-latency trade-offs across retrieval algorithms, but did not release a configurable benchmarking tool. Concurrent with our work, RAGPerf~\cite{li2026ragperf} reports system-level metrics such as throughput, memory, and utilization, but lacks support for fine-grained RAG analysis, including energy use, latency breakdowns, intermediate-stage interactions, compression rates, and pipeline configurations. It also does not capture millisecond-scale query behavior. We address this gap with a benchmark framework that treats textual RAG as a fine-grained workload, enabling detailed profiling across retrieval, compression, generation, and system execution.
\section{Conclusion}

We present a benchmarking framework for characterizing the architectural behavior of RAG pipelines on GPU systems. Our framework decomposes end-to-end RAG execution into per-stage latency, energy, compute utilization, memory bandwidth, and memory footprint metrics across five canonical pipeline configurations: naive, reranking, compression, combined, and iterative. We  publicly release our framework, as a tool for enabling reproducible and fine grained systems analysis of RAG systems and support future exploration of pipeline designs.


\section*{Acknowledgment}
\thanks{Funded in part by the National Institute of Environmental Health Sciences (NIEHS) Superfund Research Program at the National Institutes of Health (NIH) grant P42ES017198, Puerto Rico Testsite for Exploring Contamination Threats (PROTECT).}

\bibliographystyle{IEEEtran}
\bibliography{reference}

\clearpage
\section{Artifact Appendix}

\subsection{Abstract}

RAGMark is a hardware benchmarking framework for advanced Retrieval-Augmented Generation (RAG) pipelines. The submitted code artifact contains the implementation used to generate the experimental results presented in the corresponding sections of the paper.

This appendix describes how to access the code artifact, install it on supported hardware, and reproduce the reported results. To facilitate evaluation, we also provide access to a remote execution environment that allows reviewers to run the benchmark and generate the data required to reproduce three figures from the paper.
\subsection{Artifact check-list (meta-information)}

{\small
\begin{itemize}
  \item {\bf Program: } Advanced dense-retrieval RAG benchmarking tool supporting a wide range of configurations (embed, FAISS search, optional rerank/compression, LLM generation), and both standard and iterative pipelines.
  \item {\bf Hardware: } NVIDIA GPU. Paper results: 2$\times$ A100-SXM4-40GB. Artifact evaluation machine: 2$\times$ V100-PCIE-16GB. CPU/single-GPU subsets also supported.
  \item {\bf Metrics: } RAG Accuracy: (EM, F1, ROUGE-L, retrieval recall); Hardware performance (latency, TTFT, throughput, GPU power/energy, utilization).
  \item {\bf Output: } Per-query CSV/JSONL (accuracy, timing, energy, traces).
  \item {\bf Experiments: } Three reproducible experiments corresponding to paper figures: (1) reranking latency/accuracy (Fig.~4), (2) compression-rate latency (Fig.~5), (3) TTFT-vs-ROUGE trade-off (Fig.~9).
  \item {\bf Time required?: } $\sim$1 hour for all experiments and figures.
  \item {\bf Disk space required?: } $\sim$65GB total for data and code (FAISS index, cached model weights, and the Wikipedia corpus dominate). This excludes the $\sim$10GB Docker image.
  \item {\bf Publicly available?: } Yes (GitHub: \url{https://github.com/zferic/RAGMark}; Zenodo: \url{https://doi.org/10.5281/zenodo.22058207}).
  \item {\bf Badges Applied For: } Available, Reviewed, and Reproducible.
\end{itemize}
}

\subsection{Description}

\subsubsection{How to access}
\url{https://github.com/zferic/RAGMark}, archived at DOI: \url{https://doi.org/10.5281/zenodo.22058207}.

\subsubsection{Hardware dependencies}

The artifact evaluation reproduces three experiments using a machine equipped with 2$\times$ NVIDIA V100-PCIE-16GB GPUs. The corresponding paper results were collected on 2$\times$ NVIDIA A100-SXM4-40GB GPUs, which are not available for remote access.

As a result, absolute latency, energy, and memory measurements will differ from those reported in the paper. However, the experiments remain reproducible on equivalent hardware and demonstrate RAGMark's portability across hardware platforms.

\subsubsection{Software dependencies}

The provided execution environment includes all required software dependencies, including Docker, 
CUDA, Python, and the necessary machine learning libraries. No additional installation is required 
for artifact evaluation. For local installation on a new system, all Python dependencies are listed 
in \texttt{requirements.txt}, with additional setup steps described in the project README.

\subsubsection{Data sets}
The retrieval corpus consists of the 2018 Wikipedia dump ($\sim$9.2M passages). Evaluation datasets include NQ, TriviaQA, SQuAD, WebQuestions, PopQA, and HotpotQA (via FlashRAG).

\subsection{Installation}

\subsubsection{Artifact Evaluation}
We provide a remote execution environment reachable via SSH, requiring reviewers to execute only a single script. Login credentials for a shared reviewer account are provided via HotCRP,
with no personal registration required.

\begin{verbatim}
ssh <username>@129.10.225.5
cd RAGMark
bash ae_evaluation/ae_eval.sh
\end{verbatim}

\subsubsection{From Scratch}
To install RAGMark from scratch on a new system instead (not needed for AE), follow
these steps:
\begin{verbatim}
git clone https://github.com/zferic/RAGMark.git
cd RAGMark

python -m venv .venv
source .venv/bin/activate

pip install -r requirements.txt
# faiss-gpu (or faiss-cpu)
conda install -c pytorch -c nvidia \
    faiss-gpu=1.8.0

# edit config.py: HF_HOME, BASE_OUT,
# WIKI_INDEX_DIR, WIKI_CORPUS_2018, DATASET_DIR

# build/place a FAISS index (Flat/IVF/IVF-SQ) under
# WIKI_INDEX_DIR via FlashRAG, plus the corpus JSONL
# and FlashRAG eval JSONL files under DATASET_DIR
\end{verbatim}

\subsection{Experiment workflow}

The provided script, \texttt{ae\_evaluation/ae\_eval.sh}, launches the prepared Docker environment,
 executes the evaluation experiments, and runs the analysis notebook in headless mode. Upon completion, 
 all reproduced figures are written to \texttt{ae\_evaluation/figures/}.

The script executes the following experiments, followed by the analysis notebook (executed headlessly), which renders the reproduced figures directly to \texttt{ae\_evaluation/figures/}:

\begin{itemize}
\item \textbf{Case 1 -- Naive Pipeline Analysis / Trade-offs (Sections 4.2 \& 4.8, Fig.~9):} Evaluates the latency--accuracy trade-offs of the baseline RAG pipeline across multiple retrieval depths.

\item \textbf{Case 2 -- Reranking Analysis (Section 4.3, Fig.~4):} Evaluates the latency and accuracy trade-offs of document reranking across multiple retrieval depths.

\item \textbf{Case 3 -- Compression Analysis (Section 4.4, Fig.~5):} Evaluates the impact of prompt compression on retrieval and generation performance.
\end{itemize}

\subsection{Evaluation and expected results}

Running \texttt{ae\_evaluation/ae\_eval.sh} automatically executes the evaluation experiments and the accompanying analysis notebook. Upon completion, the reproduced figures are written to \texttt{ae\_evaluation/figures/plots/} along with the aggregated CSV files used to generate each figure in \texttt{ae\_evaluation/figures/data/}.
Reviewers should expect to reproduce the following figures from the paper:
\begin{itemize}
\item Case 1 -- Naive Pipeline Analysis (Sections 4.2 \& 4.8, Fig.~9)
\item Case 2 -- Reranking Analysis (Section 4.3, Fig.~4)
\item Case 3 -- Compression Analysis (Section 4.4, Fig.~5)
\end{itemize}

\end{document}